\documentclass[draft,jgrga]{agutex}

 \usepackage{lineno}
 \linenumbers*[1]

  \usepackage[dvipdf]{graphicx}

  \setkeys{Gin}{draft=false}
\authorrunninghead{YERMOLAEV ET AL.}

\titlerunninghead{EXTREME MAGNETIC STORMS}

\begin{document}

%
%

\title{
Occurrence rate of extreme magnetic storms
}
%
%

%
%




\authors{Yu. I. Yermolaev, \altaffilmark{1}
I. G. Lodkina, \altaffilmark{1} 
N. S. Nikolaeva , \altaffilmark{1} 
M. Yu. Yermolaev \altaffilmark{1}
}

\altaffiltext{1}{Space Plasma Physics Department, Space Research Institute, 
Russian Academy of Sciences, Profsoyuznaya 84/32, Moscow 117997, Russia. 
(yermol@iki.rssi.ru)}






%
%


\begin{abstract} 

Statistical analysis of occurrence rate of magnetic storms induced 
by different types of interplanetary drivers is made on the basis of OMNI data 
for period 1976-2000. 
Using our catalog of large scale types of solar wind streams we study storms 
induced by interplanetary coronal mass ejections (ICME) (separately magnetic clouds (MC) 
and Ejecta) and both types of compressed regions: 
corotating interaction regions (CIR) and Sheaths. 
For these types of drivers we calculate integral probabilities of storms 
with minimum $Dst \le$ -50, -70, -100, -150, and -200 nT. 
The highest probability in this interval of Dst  is observed for MC, 
probabilities for other drivers are 3-10 times lower than for MC. 
Extrapolation of obtained results to extreme storms shows that such a magnetic storm as  
Carrington storm in 1859 with $Dst =$ -1760 nT is observed on the Earth with frequency 1 event
during $\sim$ 500 year.

\end{abstract}

%
%

%

\begin{article}

%
%

\section{Introduction}

One of the main problems of solar-terrestrial physics and Space Weather is study and 
prediction of magnetic storms  including extreme ones 
(see, e.g., recent papers and reviews by 
\cite{Echeretal2011,PodladchikovaPetrukovich2012,Yakovchouketal2012,Yermolaevetal2012}
and references therein). 
Useful technique for problem solution is calculation of frequency (occurrence rate) distributions of events 
based on observations in form $dN =F(x)dx$, where $dN$ is the number of events recorded 
with the parameter $x$ of interest between $x$ and $x+dx$, and $F(x)$ is a frequency distributions 
(see, e.g., recent review  and paper by 
\cite{Crosby2011,GorobetsMesserotti2012}
and references therein). 
The $Dst$ index (or its proxies) calculated from the observations of the horizontal 
magnetic field at four low- to mid-latitude ground stations is used to indicate value of magnetic storms 
\citep{Sugiura1964, SugiuraKamei1991}. 
Recently the geomagnetic storm peak intensity distribution has been obtained for period 1957 to 2008 and 
it has been described by an exponential form, $F(Dst_p) =1.2 exp(-Dst_p/34)$ 
where $F$ is the probability of geomagnetic storm occurrence with a given value $Dst_p$ 
\citep{Echeretal2011}. 
This formula gives too small probability for extreme storms 
(for instance, for storms such as the 1859 Carrington storm with $Dst = - 1760$ nT  
\citep{Tsurutanietal2003} 
probability is equal to 1 event per 10$^{23}$ magnetic storms 
\citep{Echeretal2011}
). 

To predict time and value of magnetic storm it is necessary to additionally find relationships 
between interplanetary and magnetospheric parameters/events. 
\cite{Alvesetal2011} 
analyzed interplanetary structures (magnetic clouds with sheath before them -- MC, 
corotating interaction regions -- CIR, and shocked plasma) and distributions 
(histograms and fitting lines) for geomagnetic indices for each interplanetary structure 
(the geomagnetic indices $Kp (ap)$, $AE$, and $Dst$ are peak values within 2 days after 
the interplanetary structure has passed near- Earth orbit). 
The main disadvantages and limitations of this work are following.

1. MC, Ejecta and Sheath events are not selected and analyzed together as "MC events" 
though it is well known that these types of interplanetary drivers result in different 
reaction of magnetosphere 
\citep{TsurutaniGonzalez1997,Gonzalezetal1999,Yermolaevetal2005,WuLepping2007,Pulkkinenetal2007,Yermolaevetal2010c,Kilpuaetal2012}

2. The used technique of relating of interplanetary driver to magnetic storm is ambiguous 
because the delay between the reason and its consequence usually does not exceed 2 hours
\cite{GonzalezEcher2005}, 
and usually duration of interplanetary drivers (average durations 24 $\pm$ 11 h for
MC, 29 $\pm$ 5 h for Ejecta, 16 $\pm$ 3 h for Sheath before Ejecta, 9 $\pm$ 5 h for Sheath before MC, 
and 20 $\pm$ 4 h for CIR 
\citep{Yermolaevetal2007c,Yermolaevetal2010a}) 
is less than 2 days. 

3. Used approximations at large indices (see Table 4 in paper by 
\cite{Alvesetal2011})
give constant values 0.06, 0.22 and 0.045 for MC, shock and CIR, respectively, 
and cannot be used for analysis of extreme storms.

In our previous paper 
\citep{Yermolaevetal2012} 
we analyzed 798 geomagnetic storms with $Dst \le -50$ nT and five various types of solar wind 
streams as their interplanetary sources: corotating interaction regions (CIR), 
interplanetary coronal mass ejection (ICME) including magnetic clouds (MC) and ejecta, and 
a compression region sheath before both types of ICME ($SH_{MC}$ and $SH_{Ej}$, respectively) 
and calculated a probability with which a selected phenomenon can cause a magnetic storm, 
i.e., the ratio between the number of events $K_j$ of a chosen stream type $j$ (MC, CIR etc.) 
resulting in a magnetic storm with $Dst < Dst_0$ and the total number of this type events 
$N_j$: $P_j = K_j/N_j$. 
In contrast with papers by 
\cite{Alvesetal2011} and 
\cite{Echeretal2011} 
we calculated an integral probability summing probabilities from $-\infty$ up to $Dst_0$, 
i.e.  $P_j(Dst_0) = \int_{-\infty}^{Dst_0} F_j(Dst) d(Dst)$. 
In previous paper we calculated $P(-50)$ for MC, Ejecta, Sheath and CIR. 
In this paper we calculate integral probabilities for $|Dst| = 50, 70, 100, 150, 200$ nT and 
extrapolate them to stronger, extreme storms with $|Dst| = 500, 1000, 1700$ nT.

\section{Methods} 

We use our catalog of large-scale solar wind phenomena in 1976--2000 (see ftp://www.iki.rssi.ru/pub/omni  
\citep{Yermolaevetal2009} 
obtained on the basis of OMNI database (see http://omniweb.gsfc.nasa.gov 
\citep{KingPapitashvili2004} 
and data on $Dst$ index (see http://wdc.kugi.kyoto-u.ac.jp/index.html). 
The technique of determination of connection between magnetic storms and 
their interplanetary drivers consists in the following. 
If the minimum of $Dst$ index lies in an interval of a type of solar wind streams or 
is observed within 1--2 hours after it we believe that the given storm has been generated 
by the given type of streams 
\citep{Yermolaevetal2010}. 
During 1976-200 there were 798 magnetic storms with minimum $Dst \le $ -50 nT 
but the source of 334 magnetic storms (i.e., 42 \% of 798 storms) 
occurred to be undetermined, mainly because of the absence of a complete
set of measurements for separate time intervals in the OMNI database.

We use indicated data to calculate following parameters: (1) integral probability $P_j(Dst_0)$ 
and (2) waiting time $T_j(Dst_0)$. 
$P_j(Dst_0) = K_j(Dst_0)/N_j$, 
where $N_j$ is total number of solar wind events of type $j$ observed during period 1976-2000, 
$K_j(Dst_0)$ is number of solar wind events of type $j$ induced magnetic storms 
with minimum $Dst \le Dst_0$, 
i.e. $P_j(Dst_0)$ is function of threshold $Dst_0$. In this work we calculate integral probability for 
values $Dst_0$ = -50, -70, -100, -150, and -200 nT and for 4 types of interplanetary drivers 
( CIR, Sheath, MC, Ejecta as well as ME=MC+Ejecta) on the basis of experimental data. 
These experimental points are approximated by different functions and obtained approximations allows us to 
extrapolate probabilities to extreme magnetic storms with high values $Dst_0$ = -500, -1000, and -1700 nT 
for which experimental data are too scare to calculate real probabilities.   

An important parameter of magnetic storms is a waiting time which is average period 
between consecutive observations of magnetic storms stronger $Dst_0$ generated by solar wind type $j$, i.e.  
$T_j(Dst_0) = [P_j(Dst_0)N_j/(t * r)]^{-1}$, where $t$ = 26 years 
(duration of observations from 1976 up to 2000) 
and $r$ = 0.58 (ratio of durations of data existence to total duration of observation). 
To estimate waiting times $T_j(Dst_0)$ for $Dst_0$ = -500, -1000, and -1700 nT, 
we use approximating functions pf integral probabilities at these values of $Dst$ index.      

\section{Results}

Total numbers of different interplanetary drivers $N_j$, 
numbers of drivers resulting in magnetic storms $K_j(Dst_0)$ and 
integral probabilities $P_j(Dst_0)$ for $|Dst_0| = 50, 70, 100, 150, 200$ nT are presented in Table 1. 
Data in last column with $K_j(-200) =$ 3 and 5 for CIR and Ejecta were excluded from consideration of $P_j$ 
because of too low statistics. Obtained data on $P_j$ are presented in Figure 1. 

These data shows that the highest probability is observed for magnetic clouds. 
At $Dst_0$ index increase from 50 to the 200 nT the probability fell almost in 10 times.
The probability for other types is less in 3-4 times at $|Dst_0|$ = 50 and 5-15 times at 200 nT, 
i.e. the probability for magnetic clouds decreases more slowly with increasing index than for other drives. 
It is important to note that ratio of probabilities of MC and Ejecta changes from 5 at $|Dst_0|$ = 50 nT 
upto 10 at $|Dst_0|$ = 150 nT and probability of ICME (indicated as ME = MC + Ejecta) is close to 
probability of Ejecta, and this fact is serious argument to analyze MC and Ejecta separately.

\section{Discussion} 

Obtained results allow us to calculate probabilities and waiting times of magnetic storms 
in $Dst$ range from -50 down to - 200 nT for various interplanetary drivers and 
to use these parameters for forecasting of magnetic storms. Unfortunately, there is no enough number 
of events to calculate these parameters for larger storms. 
Nevertheless obtained data can be approximated by functions and 
these approximations can be used for extrapolation of 
probabilities in range of very large storms. 

In accordance with paper by 
\cite{Crosby2011} 
various events on the Sun and the Earth have similar shape of distributions: 
a top at small values, then a gentle slop and power--law tail at large values. 
If suggestion on power--law distribution at given values of $Dst$ is true, 
obtained data can be approximated by power--law functions and these approximations were 
made using last highest 3 measured points of $|Dst|$ (see Table 2 and thing lines in figure 2). 
As seen in Fig.2, 
dependence of $log(P)$ on  $log(|Dst|)$ is not linear (as it should be for power law). 
So we made square approximation in log-log scales (see Table 2 and thick lines in figure 2). 
The square-law dependencies are too steeply decreasing and do not agree conclusion by 
\cite{Crosby2011} about power--low tail at large values. 
We suggest that dependencies do not approach power--law part of distribution 
in range of $|Dst_0| = 50, 70, 100, 150, 200$. 
So, we take power law with fixed index -2.5 for MC (see Table 2 and dashed line in figure 2)
and believe that this line is the best approximation. The probabilities for another drivers 
in these range of $Dst$ have been approximated by power law with fixed index -2.5 (not shown) 
and obtained lines look better than power--law fittings on last 3 points and square--law fittings. 

Probabilities obtained by extrapolations of fittings are used for calculation of waiting times 
for extreme storms with minimum $Dst \le$ -500, -1000, and -1700 nT (see Table 2). 
The data spread is very large but it is possible to make some rough estimations. 
The least waiting times (e.g. the highest probability) is observed for MC. 
Other drivers have larger waiting times (lower probabilities) 
and their contribution can be neglected. 
Thus, the most probable estimations of waiting times for extreme magnetic storms 
with minimum $Dst \le$ -500, -1000, and -1700 nT are $\sim$20, $\sim$100 and $\sim$500 years 
(with accuracy of factors $\sim$1.5, $\sim$2, and $\sim$3), respectively.  
It should be noted that we use 26 year observations to estimate waiting time of 100 and 500 years, 
therefore accuracy of estimates is very low. 
Nevertheless the obtained estimates appear more realistic than ones were obtained by 
\cite{Echeretal2011}. 
In order essentially to increase accuracy and reliability of estimates by means of this technique 
it is necessary to increase duration of an interval of observation several times.

\section{Conclusions} 

On the basis of OMNI dataset during 1976-2000 we classified different types of solar wind events 
(magnetic cloud MC, Ejecta, CIR, and Sheath), 
found magnetic storms corresponding to these types of interplanetary drivers and calculated 
integral probabilities of generation of magnetic storms with minimum $Dst \le$ -50, -70, 
-100, -150 and -200 nT by these types of solar wind drivers. 
Obtained data allow one to make following conclusions.

1. Probabilities of storm generation by all types of drivers decrease with decreasing minimum $Dst $ index, 
and rate of probability decrease increases with index decreasing.

2. Probability for MC is the highest one, and rate of its probability decrease is the smallest. 

3. Probabilities of CIR--, Ejecta-- and Sheath--induced storms with $Dst \le $ -50 nT 
are less in 3-4 times than for MC, for storms with $Dst \le $ -200 nT their probabilities 
are less in 5-15 times.

Interpolation of data in $Dst$ region from -50 down to -200 nT and than extrapolation 
for stronger, extreme storms with $Dst \le $ -500, -1000, and -1700 nT allow us to suggest the following.   

1. The most probable estimations of waiting times for extreme magnetic storms 
with minimum $Dst \le$ -500, -1000, and -1700 nT are $\sim$20, $\sim$100 and $\sim$500 years 
(with accuracy of factors $\sim$1.5, $\sim$2, and $\sim$3), respectively.  

2. The waiting times of CIR--, Ejecta-- and Sheath--induced storms with $Dst \le $ -500 nT 
are larger in 5-10 times than for MC, their waiting times for storms with $Dst \le $ -1700 nT 
are larger in 10-100 times than for MC. Therefore their waiting times for extreme storms is very
small relative to MC one and their contribution in extreme storm generation can be neglected. 

Obtained experimental results and estimations are important for space weather 
and, in particular, for analysis and forecasting of extreme magnetic storms.


%
%
%
%
%
%
%

\begin{acknowledgments}
The authors are grateful for the opportunity to use the OMNI database. The OMNI data were obtained from 
GSFC/ SPDF OMNIWeb (http://omniweb.gsfc.nasa.gov). This work was supported by the 
Russian Foundation for Basic Research, projects 10–02–00277a and 13-02-00158, and by 
Program 22 of Presidium of the Russian Academy of Sciences.
\end{acknowledgments}

\end{article}


%
%

%
%
%
%
%


\begin{figure}
\noindent\includegraphics[width=10cm]{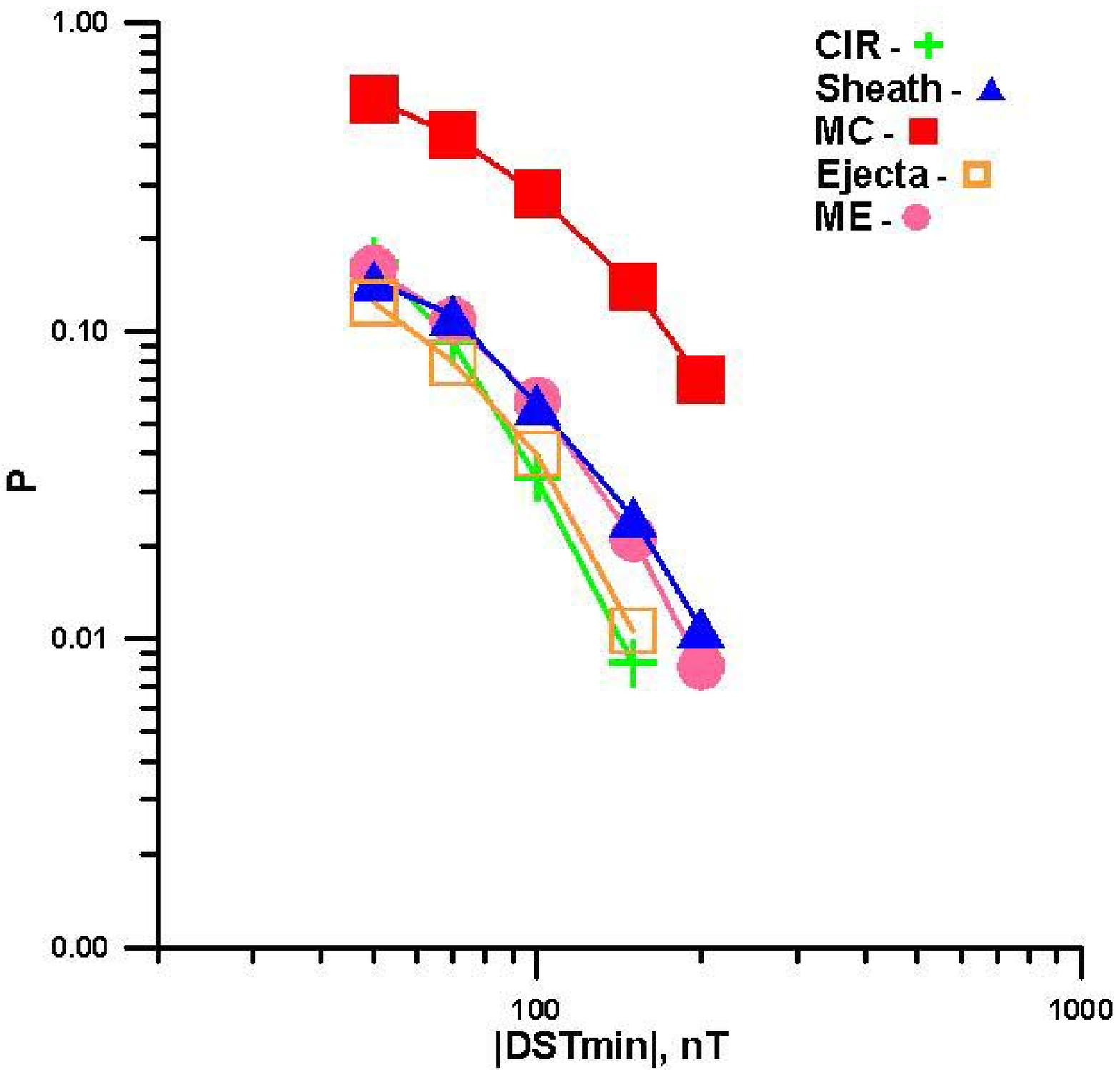}
\caption{Dependence of integral probability $P_j(|Dst|> Dst_0)$ on the value of storm for different types of solar wind.}
\end{figure}


\begin{figure}
\noindent\includegraphics[width=10cm]{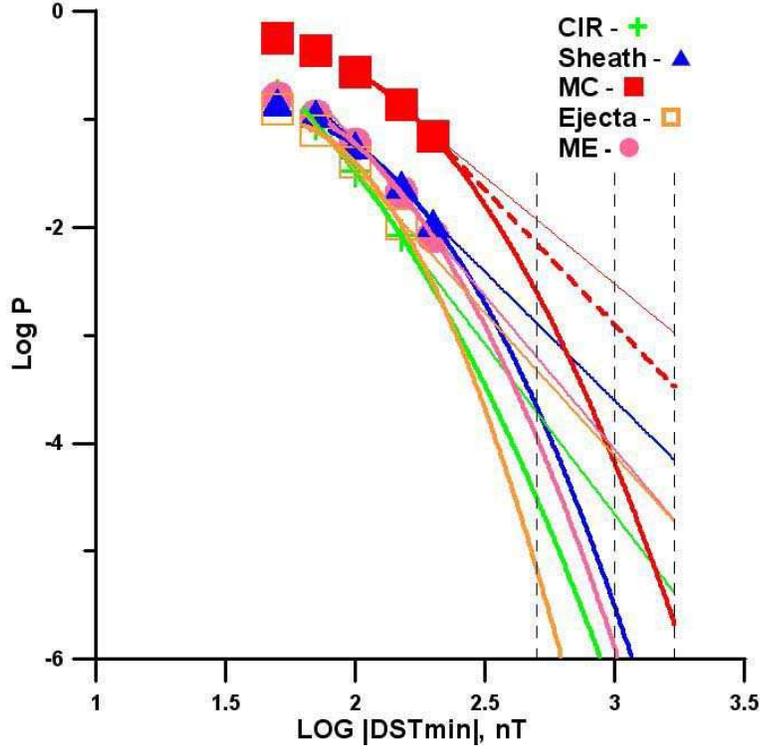}
\caption{Approximations of dependence of integral probability $P(|Dst|> Dst0)$ on the value of storm for different types of solar wind. 
Vertical dashed lines show 500, 1000 and 1700 nT.}
\end{figure}

%

\begin{table}
\caption{Probabilities $P_j$ generation of magnetic storms with $Dst \le$ -50, -70, -100, -150 and -200 nT for different types of interplanetary drivers.
}
\centering
\begin{tabular}{l|l|cc|cc|cc|cc|cc}
\hline
Types of & Number, & \multicolumn{2}{c}{$Dst \le -50$} & 
               \multicolumn{2}{|c}{$Dst \le -70$}  &
	\multicolumn{2}{|c}{$Dst \le -100$}  &
	\multicolumn{2}{|c}{$Dst \le -150$}  &
	\multicolumn{2}{|c}{$Dst \le -200$}  \\
\cline{3-12}
  drivers &  $N_j$ & $K_j$ & $P_j$ & $K_j$ & $P_j$ & $K_j$ & $P_j$ & $K_j$ & $P_j$ & $K_j$ & $P_j$ \\
\hline
  CIR                 & 718 & 120  & 0.167 & 66 & 9.2  $10^{-2}$ & 24 & 3.3 $10^{-2}$ & 6 & 8.4 $10^{-3}$ & 5 & - \\
  Sheath              & 642 &  93  & 0.145 & 72 & 0.112 & 37 & 5.8 $10^{-2}$ & 16 &2.6 $10^{-2}$ & 7 & 1.1 $10^{-2}$ \\
  MC                  & 101 &	57 & 0.564 & 44 & 0.436 & 28 & 0.227 & 14 & 0.139 & 7 & 6.9 $10^{-2}$ \\
  Ejecta              & 1127 & 139 & 0.123 & 89 & 7.9 $10^{-2}$ & 45 & 4.0 $10^{-2}$ & 12 & 1.1 $10^{-2}$ & 3 & - \\
  ME (MC + Ejecta)    & 1228 & 196 & 0.159 & 133& 0.108 & 73 & 5.9 $10^{-2}$ & 26 & 2.1 $10^{-2}$ & 10 & 8.1 $10^{-3}$  \\
\hline
\end{tabular}
\end{table}

\begin{table}
\caption{Approximations of integral probabilities $P_j$ and waiting times (average occurrence periods) $T_j$ 
for magnetic storms with $|Dst| > 500, 1000, 1700$ nT.
} 
\centering
\begin{tabular}{l|l|cc|cc|cc}
\hline
Types of &  Extrapolation,  & \multicolumn{2}{c}{$ |Dst| > 500 $} & 
	                      \multicolumn{2}{|c}{$ |Dst| > 1000 $} &
                              \multicolumn{2}{|c}{$ |Dst| > 1700 $}  \\
\cline{3-8}
 drivers &  $Y = log(P), X = log(|Dst|)$& $P$ & $T$ & $P$ & $T$ & $P$ & $T$  \\
\hline
MC	&$Y= -1.98 X + 7.87 	$	&11.18 $10^{-2}$&12.7	&2.99 $10^{-3}$	&50	&1.05 $10^{-3}$	&142 \\
	&$Y=-2.5 X+4.6		$	&7.12 $10^{-3}$	&21.0	&1.26 $10^{-3}$	&118	&3.34 $10^{-4}$	&447 \\
	&$Y = -4.19 + 5.14 X - 1.66 X^2$&3.87 $10^{-3}$	&38.6	&1.93 $10^{-4}$	&774	&$1.22 10^{-5}$	&12200 \\
\hline
Ejecta	&$Y= -3.26 X + 5.12 $		&2.10 $10^{-4}$	&63.7	&2.20 $10^{-5}$	&608	&$3.89 10^{-6}$	&3400 \\
	&$Y = -8.73 + 9.91X - 3.13 X^2$	&1.74 $10^{-5}$	&770	&7.23 $10^{-8}$	&1.9 $10^5$	&$4.49 10^{-10}$&3.0 $10^7$ \\
\hline
ME	&$Y = -2.85 X + 4.48 	$	&6.31 $10^{-4}$	&19.5	&8.77 $10^{-5}$	&140	&$1.94 10^{-5}$	&630 \\
	&$Y= -6.39+ 7.30 X - 2.36 X^2$	&1.34 $10^{-4}$ &91.6	&1.90 $10^{-6}$	&6500	&$3.75 10^{-8}$	&3.3 $10^5$ \\
\hline
CIR	&$Y = -3.42 X + 5.36 	$	&1.36 $10^{-4}$	&154	&1.27 $10^{-5}$	&1650	&$2.08 10^{-6}$	&10100 \\
	&$Y = -5.51 + 7.11 X - 2.54 X^2$&1.42 $10^{-5}$	&1480	&8.53 $10^{-8}$	&2.5 $10^{5}$ &8.28 $10^{-10}$ &2.5 $10^{7}$ \\
\hline
Sheath	&$Y = -2.38 X + 3.54 	$	&1.30 $10^{-3}$	&18.1	&2.49 $10^{-4}$	&94.3	&7.05 $10^{-5}$	&333 \\
	&$Y = -5.35 + 6.01 X - 1.97 X^2$&3.19 $10^{-4}$	&73.6	&8.50 $10^{-6}$	&2760	&3,03 $10^{-7}$	&7.8 $10^{4}$ \\
\hline
\end{tabular}
\end{table}


\begin{thebibliography}{}



\bibitem[{\textit{Alves et al.}(2011)}]{Alvesetal2011}
Alves M.V., E. Echer, W.D. Gonzalez, Geoeffectiveness of solar wind interplanetary magnetic structures, 
Journal of Atmospheric and Solar-Terrestrial Physics 73 (2011) 1380-–1384 , doi:10.1016/j.jastp.2010.07.024  

\bibitem[{\textit{Crosby}(2011)}]{Crosby2011}
Crosby, N. B.. Frequency distributions: from the sun to the earth, Nonlinear Processes in Geophysics, 
18, 791-–805, 2011, doi:10.5194/npg-18-791-2011

\bibitem[{\textit{Echer et al.}(2011)}]{Echeretal2011}
Echer E.,W.D. Gonzalez and B.T.Tsurutani, Statistical studies of geomagnetic storms with peak 
$Dst \le -50$ nT from 1957 to 2008, 
Journal of Atmospheric and Solar–Terrestrial Physics, 73 (2011) 1454-–1459, doi:10.1016/j.jastp.2011.04.021 

\bibitem[{\textit{Gonzalez et al.}(1999)}]{Gonzalezetal1999}
Gonzalez, W. D., Tsurutani, B. T., and Clua de Gonzalez, A. L. (1999), 
Interplanetary origion of geomagnetic storms, 
{\it Space Sci. Rev.,} \textit{88}, 529--562.

\bibitem[{\textit{Gonzalez and Echer}(2005)}]{GonzalezEcher2005}
Gonzalez, W.D., and Echer, E., A (2005), 
Study on the Peak Dst and Peak Negative Bz Relationship during Intense 
Geomagnetic Storms, {\it Geophys. Res. Lett.,}  \textit{32}, L18103. 
doi: 10.1029/2005GL023486.

\bibitem[{\textit{Gorobets and Messerotti}(2012)}]{GorobetsMesserotti2012}
Gorobets A., M. Messerotti, Solar Flare Occurrence Rate and Waiting Time Statistics, Solar Phys (2012), 
 281:651–667, DOI 10.1007/s11207-012-0121-7 

\bibitem[{\textit{Kilpua et al.}(2012)}]{Kilpuaetal2012}
Kilpua E. K. J., L. K. Jian, Y. Li, J. G. Luhmann, C. T. Russell, 
Observations of ICMEs and ICME-like Solar Wind Structures from 2007–-2010 
Using Near-Earth and STEREO Observations
Solar Physics (2012), Volume 281, Issue 1, pp 391-409 

\bibitem[{\textit{King and Papitashvili}(2004)}]{KingPapitashvili2004}
King, J.H. and Papitashvili, N.E., (2004), 
Solar Wind Spatial Scales in and Comparisons of Hourly Wind and ACE Plasma and 
Magnetic Field Data, J. Geophys. Res.,vol. 110, no. A2, p. A02209. doi: 10.1029/2004JA010804. 

\bibitem[{\textit{Podladchikova and Petrukovich}(2012)}]{PodladchikovaPetrukovich2012}
Podladchikova, T. V., and A. A. Petrukovich (2012), Extended geomagnetic storm forecast 
ahead of available solar wind measurements, 
Space Weather, 10, S07001, doi:10.1029/2012SW000786. 

\bibitem[{\textit{Pulkkinen et al.}(2007)}]{Pulkkinenetal2007}
Pulkkinen, T. I., Partamies, N., Huttunen, K. E. J., Reeves, G.
D., and Koskinen, H. E. J.: Differences in geomagnetic storms
driven by magnetic clouds and ICME sheath regions, Geophys.
Res. Lett., 34, L02105, doi:10.1029/2006GL027775, 2007

\bibitem[{\textit{Sugiura}(1964)}]{Sugiura1964}
Sugiura, M. (1964), Hourly values of equatorial $Dst$ for the IGY, Ann. Int. Geophys. Year, 35, 9.

\bibitem[{\textit{Sugiura and Kamei}(1991)}]{SugiuraKamei1991}
Sugiura, M., and T. Kamei (1991), 
Equatorial $Dst$ index 1957--1986, in IAGA Bulletin, vol. 40, edited by A. Berthelier and M. Menvielle, 
ISGI Publ. Off., Saint-Maur-des-Fosses, France 

\bibitem[{\textit{Tsurutani and Gonzalez}(1997)}]{TsurutaniGonzalez1997}
Tsurutani, B. T. and Gonzalez, W. D. (1997), 
The interplanetary Causes of Magnetic Storms: A Review, in: Magnetic Storms, 
edited by: Tsurutani, B. T., Gonzalez, W. D., and Kamide, Y., Amer. Geophys. Union Press, 
Washington D.C., Mon. Ser., 98, p. 77, 1997.

\bibitem[{\textit{Tsurutani et al.}(2003)}]{Tsurutanietal2003}
Tsurutani, B. T., W. D. Gonzalez, G. S. Lakhina, and S. Alex (2003), 
The extreme magnetic storm of 1–2 September 1859, J. Geophys. Res., 108(A7), 1268, doi:10.1029/2002JA009504. 
 
\bibitem[{\textit{Wu and Lepping}(2007)}]{WuLepping2007}
Wu, C.-C., and R. P. Lepping, Comparison of the characteristics of magnetic clouds
and magnetic cloud-like structures for the events of 1995--2003, Solar Phys., 242 ,
159–165, 2007.

\bibitem[{\textit{Yakovchouk et al.}(2012)}]{Yakovchouketal2012}
Yakovchouk, O. S., K. Mursula, L. Holappa, I. S. Veselovsky, and A. Karinen (2012), Average properties
of geomagnetic storms in 1932--2009, J. Geophys. Res., 117, A03201, doi:10.1029/2011JA017093. 

\bibitem[{\textit{Yermolaev et al.}(2005)}]{Yermolaevetal2005}
Yermolaev Yu. I., M. Yu. Yermolaev, G. N. Zastenker, L.M.Zelenyi, A.A. Petrukovich, J.-A. Sauvaud, 
Statistical studies of geomagnetic storm dependencies on solar and interplanetary events: a review, 
Planetary and Space Science, 53/1-3 pp. 189-196, 2005 

\bibitem[{\textit{Yermolaev et al.}(2007)}]{Yermolaevetal2007c} 
Yermolaev, Yu. I., Yermolaev, M. Yu., Nikolaeva, N. S., and Lodkina, L. G. (2007), 
Interplanetary conditions for CIR-induced and MC induced geomagnetic storms, Bulg. J. Phys., 34, 128--135. 

\bibitem[{\textit{Yermolaev et al.}(2009)}]{Yermolaevetal2009} 
Yermolaev, Yu. I., et al., (2009), 
Catalog of Large-Scale Solar Wind Phenomena during 1976--2000, 
Kosm. Issled.,  vol. 47, no. 2, pp. 99--113. [Cosmic Research, pp. 81--94].

\bibitem[{\textit{Yermolaev et al.}(2010c)}]{Yermolaevetal2010c} 
Yermolaev, Yu. I., N. S. Nikolaeva, I. G. Lodkina, and M. Yu. Yermolaev, (2010c), 
Specific interplanetary conditions for CIR-, Sheath-, and ICME-induced geomagnetic storms 
obtained by double superposed epoch analysis, {\it Ann. Geophys.,}  \textit{28}, 2177--2186.

\bibitem[{\textit{Yermolaev et al.}(2010)}]{Yermolaevetal2010} 
Yermolaev, Yu. I., N. S. Nikolaeva, I. G. Lodkina, and M. Yu. Yermolaev, (2010), 
Relative Occurrence Rate and Geoeffectiveness of 
Large-Scale Types of the Solar Wind, Kosmicheskie Issledovaniya, Vol. 48, No. 1, pp. 3--32. 
(Cos. Res., 2010, Vol. 48, No. 1, pp. 1--30).

\bibitem[{\textit{Yermolaev et al.}(2010a)}]{Yermolaevetal2010a} 
Yermolaev, Yu. I., N. S. Nikolaeva, I. G. Lodkina, and M. Yu. Yermolaev, (2010a), 
Relative Occurrence Rate and Geoeffectiveness of 
Large-Scale Types of the Solar Wind, Kosmicheskie Issledovaniya, Vol. 48, No. 1, pp. 3--32. 
(Cos. Res., 2010, Vol. 48, No. 1, pp. 1--30).

\bibitem[{\textit{Yermolaev et al.}(2012)}]{Yermolaevetal2012}
Yermolaev, Y. I., N. S. Nikolaeva, I. G. Lodkina, and M. Y. Yermolaev (2012), 
Geoeffectiveness and efficiency of CIR, sheath, and ICME 
in generation of magnetic storms, J. Geophys. Res., 117, A00L07, doi:10.1029/2011JA017139





\end{thebibliography}
\end{document}